# A conformal mapping approach to broadband nonlinear optics on chip

Chunyu Huang[1], Yu Luo[2,3,*], Yule Zhao[1], Xiaofei Ma[1], Zhiwei Yan[1], Ziyi Liu[1], Chong Sheng[1], Shining Zhu[1], Hui Liu[1,*]

[1]Collaborative Innovation Center of Advanced Microstructures National Laboratory of Solid State Microstructures，School of Physics Nanjing University Nanjing Jiangsu 210093 China

[2]School of Electrical and Electronic Engineering, Nanyang Technological University, 50 Nanyang Avenue, Singapore, 639798, Singapore

[3]Key Laboratory of Radar Imaging and Microwave Photonics, Nanjing University of Aeronautics and Astronautics, Nanjing 211106, China

[*]Correspondence authors: liuhui@nju.edu.cn; luoyu@ntu.edu.sg

## Abstract

**Integrated nonlinear optical devices play an important role in modern optical communications. However, conventional on-chip optical devices with homogeneous or periodic translation dimensions generally have limited bandwidth when applied to nonlinear optical applications. Up today, there lacks a general method to design compact nonlinear optical devices over a broadband continuous frequency range. In this work, we propose a general strategy based on transformation optics (TO) to design curved accelerating waveguides (CAWs) with spatially gradient curvatures able to achieve broadband nonlinear frequency conversion on chip. Through rigorous analytical calculation, we show that increasing the acceleration (i.e. gradient in the waveguide curvature) broadens the output signal spectrum in the nonlinear process. In the experiment, we take the sum-frequency generation for infrared signal upconversion (SFG-ISU) as the example and fabricated a variety of CAWs using thin-film lithium niobate on insulator (LNOI). Efficient SFG is observed over a broadband continuous spectrum. Our conformal mapping approach offers a platform for various nonlinear optical processes and works in any frequency range, including visible, infrared and terahertz bands. Apart from LNOI, our approach is also compatible with other nonlinear materials, such as silicon, silicon nitride and chalcogenide glasses etc.**

Broadband nonlinear optical structures have versatile applications in high-speed optical communication and computing [1,2,3,4], integrated quantum chips[5], long-wave photon detection[6], material spectral analysis[7], microwave photonics[8,9], broadband spectrum detection[10], molecular sensing and spectroscopy[11,12,13], etc. The high demand for integrated optical applications[14], including transmitters, modulators[15,16], wavelength division multiplexers[17], call for urgent needs of broadband nonlinear optical components on chip. To this end, various approaches have been proposed, for example, in the domain of third-order nonlinearities($\chi^{(3)}$), group-velocity dispersion (GVD) engineering has been applied to broadband four-wave-mixing[18] and cascaded waveguides have enabled supercontinuum light generation[19,20]. Through GVD tuning, broadband nonlinear effects related to quadratic nonlinearities ($\chi^{(2)}$) have also been observed[21], and entangled photons were generated in periodically poled lithium niobate (PPLN) waveguides[22]. Owing to the chromatic dispersion effect in nonlinear materials, however, these nonlinear processes generally have limited bandwidths. Although the frequency comb of integrated optical microcavity offers a possible route to design multi-frequency light sources[23, 24], the output signal is not strictly continuous over a broadband spectrum.

In this work, we deploy conformal transformation optics (CTO) to break the conservation law of photon momentum, so as to obtain broadband nonlinear optical responses over a continuous spectrum. This approach is highly suitable to design translational/rotational symmetry-broken structures, whose nonlinear responses are difficult to investigate with conventional approach[18, 21, 22]. Moreover, it is not necessarily limited to $\chi^{(2)}$ or $\chi^{(3)}$, and is applicable to a general class of nonlinear processes. Last but not least, our experimental realizations also extend the domain of TO from linear to nonlinear optics. TO relates a novel optical phenomenon (which is in general difficult to achieve with traditional optical methods) to the material parameters required for its occurrence[25, 26]. It provides additional degree of freedom to manipulate electromagnetic waves through properly engineering the permittivity and permeability across the space and time[27, 28]. TO has various novel applications, including cloaking[29, 30, 31, 32, 33, 34, 35], illusion[36] and analogue of cosmic phenomenons[37, 38, 39, 40, 41, 42]. However, since materials obtained from TO are often strongly inhomogeneous and extremely anisotropic, these applications are generally difficult to realize in experiments, especially at high frequencies (e.g. infrared, visible, and beyond). Conformal and quasiconformal mappings solve this problem by eliminating the extreme anisotropy[26, 43, 44, 45], pushing the TO technology one step closer to real-world applications. Apart from devices mentioned above, constant curved microstuctures[46, 47, 48], broadband light harvesters[49], Luneburg lens[50] and Maxwell's fish-eye lens[51] have also been designed and implemented under CTO framework. In all relevant experimental realizations, CTO is applied mainly to linear material parameters. In this work, we go one step beyond and show that CTO offers great potentials to extend the bandwidth of nonlinear optical devices. More specifically, a general conformal mapping approach is developed to transform a CAW with a gradient radius of curvature into a homogeneous waveguide with spatially inhomogeneous refractive index. Through transformed wave-equation and effective nonlinear coefficients, we construct the nonlinear couple-wave equations (CWEs) of CAWs. Nobly, the bandwidth of the nonlinear frequency conversion can be effectively controlled by the acceleration parameter, i.e. the gradient of the waveguide curvature. The greater the acceleration, the broader the bandwidth of the output signal is. As an experimental proof, we take the SFG-ISU as the example, fabricate CAWs on LNOI and demonstrate that the SFG-ISU bandwidth can be efficiently broaden over an ultra-broad frequency range (i.e. 4 times broader as compared to a straight optical waveguide designed with the traditional phase matching approach). We highlight that the broadband SFG-ISU demonstrated here is just one specific application. Our method can be easily extended to other nonlinear processes and is compatible with other nonlinear materials. Moreover, CAW structures experimentally realized here are easy to integrate on chip, robust to the temperature, and compatible with the CMOS process.

## Results

1. **The Conformal Transformation Approach to design CAWs**

Our on-chip broadband nonlinear processes utilize CTO to design CAWs as depicted in Fig.1a, which are uniform in width but gradient in bending curvature. The cross-section is detailed in the bottom right inset, and the bandwidth broadening is illustrated in the upper right inset in Fig.1a. Such a waveguide has a width of 1um and a total length of 60um, etched in a 370nm-thick z-cut LNOI substrate. Since we employed a rectangular waveguide, the conformal mapping in each $(x, y)$ plane is same ($z = w$). Therefore, we aim to find a general 2-dimentional conformal mapping $\xi = f(\zeta)$, that transforms a curved waveguide in the physical space $\xi = x + iy$ to a straight one in the virtual space $\zeta = u + iv$. The former has a spatially gradient radius $r(v)$ being an arbitrary function of $v$. The conformal mapping $f(\zeta)$ under a given $r(v)$ ensures that the total length of the waveguide remaining unchanged under the transformation, i.e. $|d\xi| = r(v)d\theta = dv$, where $\theta(v) = \int_0^v dv'/r(v')$ denotes the angle between the tangential direction of the curved waveguide and the vertical axis of the physical coordinate. The conformal mapping is then obtained through the integration of $\theta(-i\zeta')$ in terms of $\xi = \int_0^{-i\zeta} e^{i\theta(-i\zeta')} d\zeta'$ (details derivations are provided in the Supplementary Information (S.I.) section 1.1[52]). The theory of CTO enables us to calculate the refractive index of the straight waveguide

$$n = n_\xi \left|\frac{d\xi}{d\zeta}\right| = n_\xi \left|e^{i\theta(-i\zeta)}\right| \tag{1}$$

Which is directly related to $\theta(-i\zeta)$ where $n_\xi$ is the refractive index of the material in the real space. When the radius is a constant $r_0$, $\theta(-i\zeta) = -i\zeta/r_0$ and the refractive index of the straight waveguide takes an exponential form $n = n_0 e^{u/r_0}$, in agreement with the previous study of uniformly curved waveguides[46, 52]. To design the CAW, we consider a spatially gradient radius of curvature given by,

$$r(v) = \frac{r_0'}{\ln(n_0'\sqrt{1 - av} - n_1')} \tag{2}$$

where $r_0' = 25.56um$, $n_0' = 1847.7$, and $n_1' = 1849.2$ are constants to be determined through initial conditions, $a$ is the acceleration parameter to characterize different CAWs. (In S.I., we also give transformations of another two curved waveguides to illustrate the generality).

Using the conformal mapping approach described above, the CAW in Fig.1a is transformed into a straight waveguide with a spatially gradient refractive index depicted in Fig.1b. The 2-dimentioanl index distribution in $(u, v)$-plane is depicted in the inset of Fig.1b. The index of lithium niobate in physical space are obtained from Ref.53. To demonstrate the continuous variation of the refractive index along both $u$ and $v$, we plot $n(u,v)$ as a function of $v$ ($u$) at several different $u$ ($v$) in Fig.1c (Fig.1d). The Fig.1c illustrates that $n(u,v)$ has different monotonicity for $u = -0.25\text{um},\ 0\text{um},\ 0.25\text{um}$. The Fig.1d shows the contrast of index increasing with $v$.

We are interested in the eigenmode propagating along the $v$ direction, which need to consider the the effective mode index defined as $n_{\text{eff}} = k_v/k_0$, where $k_v$ represents the propagating constant and $k_0$ denotes the free-space wavevector. The propagating constant $k_v$ can be obtained through the simulation of the mode field distribution at a fixed $v$. Resulting from the spatially inhomogeneous refractive index distribution in the transformed space, $n_{\text{eff}}$ varies along $v$. Detailed calculations show that, for a radius given by equation (2), the mode index can be written as a simple function of $v$, i.e.

$$n_{\text{eff}}(v,\omega) = n_{\text{eff}}(0,\omega)\sqrt{1-a(\omega)v} \tag{3}$$

where we have introduced the frequency dependent acceleration $a(\omega)$. Clearly, a larger acceleration corresponds to a larger gradient in the effective mode index, which in turns compensates the phase mismatches at different frequencies, giving rise to the broadband nonlinear wave-mixing process.

As the radius decreases in a CAW, the mode field distribution tends to become distorted and accumulate energy towards right side in the waveguide, as expressed in the inset of Fig.1d. This is due to the fact that the transformed refractive index $n$ exhibits an increasing trend with $u$. The distorted field distribution has two effects: 1) it tends to decrease the effective mode area of the CAW, thereby enhancing the intensity of electric field involved in the nonlinear interaction (a smaller effective mode area induces a larger electric field intensity in the waveguide, in sec.3 of S.I.), as $r(v)$ decreases (the red line plotted in S.I. Fig.s2(a)[52]); 2) it increases the mode overlap between fundamental and higher-order modes because of the asymmetric mode field distribution with respect to $u = 0$ (demonstrated in section 1.3 in S.I.[52]). Both effects contribute to the enhancement of the effective nonlinear coefficients, as shown by the blue line in Fig.s2(a)[52].

2. **Broadband Nonlinear Up-conversion by CAWs**

In this section, we demonstrate that the CAW designed above provides effective means to broaden the nonlinear process. For the effective mode index given by equation (3), the wave equation governing the electric field in the CAW is given by,

$$\nabla^2 E - \frac{n_{\text{eff}}^2(0,\omega)(1-av)}{c^2}\frac{\partial^2 E}{\partial t^2} = 0 \tag{4}$$

The propagation solution of equation (4) takes the form of,

$$E = \text{Ai}\left[\left(\frac{n_{\text{eff}}(0,\omega)k_0}{2a}\right)^{3/2}(av-1)\right] + i\text{Bi}\left[\left(\frac{n_{\text{eff}}(0,\omega)k_0}{2a}\right)^{3/2}(av-1)\right] \tag{5}$$

where Ai(·) and Bi(·) are the Airy functions of the first and the second kind, respectively. Since we are interested in the case where the refractive index $n_{eff}(v, \omega)$ varies slowly along $v$, such a solution can be approximated as $E \sim (1-av)^{-1/4} \exp[i\phi(v,\omega)]$, where the phase $\phi(v,\omega)$ of the transformed electric field in the CAW is a quadratic function of $v$, i.e. $\phi(v,\omega) = n_{eff}(0,\omega)k_0(v - av^2/4)$.

The eigenfunctions of the electric fields and effective nonlinear coefficients $d_{eff}$ enable us to couple the fundamental and harmonic waves through the nonlinear CWEs (equation (s11) in S.I.[52]). In the traditional SFG process, the phase mismatch $\Delta\phi$ between the fundamental and sum-frequency is a constant. As a result, the efficient SFG can only be achieved at a specific frequency when the phase mismatch is strictly zero, i.e. $\Delta\phi = 0$. On the contrary, in our CAWs, the spatially gradient $\Delta\phi(v,\omega)$ suppresses the phase mismatch at different frequencies giving rise to an efficient nonlinear conversion over a broadband continuous frequency range. A larger acceleration corresponds to a larger gradient in $\Delta\phi$, and hence a broader bandwidth of the frequency conversion. Moreover, the transformed eigenmode fields of fundamental and harmonic frequencies tend to enhance their nonlinear coupling coefficients $d_{eff}$ in CAWs (equation (s12) in S.I.)[52], which further increases the nonlinear conversion efficiency.

To demonstrate the broadband nonlinear response, we compare the nonlinear frequency conversion efficiencies and bandwidths of a uniformly curved waveguide (characterized by $a = 0$) and 11 CAWs with gradually increased accelerations (i.e. $a = 1.88, 5.34, ......, 73.43 \text{m}^{-1}$). As shown in Fig.2a, a larger acceleration corresponds to a larger gradient of curvature. Consequently, the mode index $n_{eff}$ (and hence the phase mismatch $\Delta\phi$) and the effective nonlinear coefficient $d_{eff}$ changes more rapidly along $v$, as depicted in Fig.2b. As mentioned before, the rapid variation in $\Delta\phi$ suppresses the phase mismatch leading to an efficient nonlinear conversion over a broadband frequency range. The results in Fig. 2c reveal that the SFG process is significantly broadens and enhances as $a$ increases. Simultaneously, a larger effective nonlinear coefficient results in significant enhancement of the SFG efficiency. This point is confirmed by Fig. 2d, which shows that the peak SFG efficiency increases from $0.4314\%\text{W}^{-1}\text{cm}^{-2}$ to $4.8155\%\text{W}^{-1}\text{cm}^{-2}$ (corresponding to an enhancement over 10 times) when $a$ varies from 0 to $73.43\text{m}^{-1}$. For comparison, we also plot the SFG efficiency of a straight waveguide, as given by the black dashed line in Fig. 2d. Its bandwidth is considerably narrower and peak value is significantly smaller than those of CAWs. Besides, we highlight that the bandwidth of the nonlinear conversion can be further extended by increasing length of the CAW (The limitation of bandwidth increasing with acceleration is discussed in detail in the S.I.). An example of 180um long waveguides in the inset of Fig. 2d illustrates that a CAW (the red line is for $a = 73.43\text{m}^{-1}$) can boost the SFG bandwidth more than 4 times to uniform waveguides (the black line is for $a = 0$).

Before we proceed to the experimental realization, we pause to highlight the necessities of CTO in our proposal. An optical structure with breaking translational/rotational symmetry, in general, does not have a well-defined wave vector. Consequently, the phase matching condition is difficult to obtain, and with conventional approaches, one cannot tell which parameter plays an essential role to achieve broadband nonlinear effects. This problem becomes even more critical for 3D structures, on which the nonlinear numerical simulations are extremely time-consuming. The proposed CTO approach solves this problem, and its significance lies in the fact that 1) it shows how the effective acceleration should be properly designed to broaden the nonlinear optical responses, 2) it reveals the hidden symmetry in the symmetry-broken structure, enabling us to obtain the coupled-mode equations to simplify the nonlinear calculations/simulations. Hence, the CTO is not merely a means of analyzing curved waveguides, but a general approach to guide the designs of a variety of nonlinear experiments with desired bandwidths.

3. **Experimental proof for Broadband Nonlinear Up-conversion**

To investigate the broadband SFG-ISU experimentally, we fabricate the designed CAWs on LNOI chip and detect the SF efficiency at different frequencies with the combined measurements of both sCOMS camera and spectrometer. Our experimental setup, shown in Fig.3a and described in detail in the Methods. Two fundamental frequencies are a fs-laser (Spectra-physics, Mai Tai HP) to tune the frequency over a continuous spectrum ($\lambda_{FF1} \in [920nm, 1010nm]$) and a continuous-wave (CW) laser fixed at $\lambda_{FF2} = 1064nm$. The sample are pumped by $TE_0$ and $TM_0$ modes via half-wave plate (HWP). The output light is filtered and then guided into the sCMOS camera in the middle channel and the spectrometer in the bottom channel for measurements.

We firstly investigate the SFG enhancement of CAWs at the fixed wavelength $\lambda_{FF1} = 960nm$, which can be directly reflected by images captured by the camera. We judiciously adjust the measurement setup so as to keep the input condition unchanged for different samples, and then combine the images of output SFG signals from 11 different CAWs into a GIF for comparison[52]. Fig. 3b displays three out of the 11 images at $a = 1.88 m^{-1}, a = 42.96 m^{-1}$ and $a = 73.43 m^{-1}$. Apparently, CAWs with larger accelerations give rise to brighter and spatially wider output spot. The average data from multiple measurements of the 11 CAWs confirms that the SFG efficiency increases with the acceleration (as shown in Fig.3c), in excellent agreement with our theoretical prediction. This result also be verified by the spectrometer, one of the tests are displayed in Fig.4a.

To analyze more quantitatively the SFG bandwidth broadening, we scan $\lambda_{FF1}$ continuously from 920nm to 1010nm. Note that both the sCMOS camera (Fig.3b) and spectrometer (Fig.4a) alone can detect the efficiency of the frequency conversion at a fixed wavelength[53, 54]. However, neither of them can be applied to the detection of broadband SFG because the transmission of the system is strongly frequency dependent (i.e. the nonlinear conversion efficiency is no longer equivalent to the signal intensity captured by the sCMOS camera or the spectrometer). To solve this problem, we introduce a fiber-lens system (FLS) to quantify firstly the transmission of the system through the second-harmonic generation (SHG) measured by our sCMOS camera/spectrometer combined system (see specific steps in the S.I. (section 2.2) for details). The broadband SFG efficiency is then obtained by normalizing the results obtained by the spectrometer with respect to the transmission spectrum. As shown in Fig.4b, the SFG spectra obtained in this way agrees quite well with the theoretically predicted ones. More interestingly, the full width at half maximum (FWHM) plotted in Fig.4c reveals that the SFG bandwidth increases dramatically with the acceleration.

## Discussion

We have demonstrated a novel physical mechanism to achieve broadband nonlinear optical process through the introducing of CTO. We especially designed and fabricated a type of CAWs to demonstrate the effectiveness of this mechanism experimentally. Our experiments yielded efficient SFG-ISU with a broadband spectrum. Notably, by adjusting the structure parameters of the waveguide, we can tune the bandwidth of the output SFG-ISU. Our results show that the bandwidth of SFG-ISU increases with the acceleration parameter, with no sacrificing of the efficiency for compressed mode field.

This work demonstrates the feasibility of broadband nonlinear devices by near-infrared up-conversion detection, which could be extended to any other wavelength band. Our method not only overcomes the limitation of materials but also provides flexible and expansible applications to many other nonlinear optical processes in integrated systems, including differential frequency generation (DFG), four-wave mixing (FWM), parametric down-conversion, etc. Moreover, the broadband nonlinear process is achieved by controlling the waveguide's geometrical shape, and the fabrication process is straightforward, which makes our strategy suitable for large-scale industrial production. In the future, our method has great potential in integrated nonlinear photonics and various related optical applications.

## Methods

**Theoretical methods.** We propose a novel conformal mapping approach to obtain the equivalent straight waveguide from a LNOI-CAW with varying radius. We substitute the transformed permittivity and permeability into the COMSOL Multiphysics to calculate the properties of CAWs, including effective mode index and mode area. Based on transformed wave function and transformed eigenmode field, we establish the nonlinear CWEs model in CAWs with varying index. Additionally, we, we numerically solved out the SFG intensity for different accelerations under the same initial conditions.

**Sample design and fabrication.** We fabricate CAWs on LNOI with varying radius as equation (2) expressed. All CAWs are defined with an initial radius of $r(0) = 60um$ and ended with $r(L) = 50, 40, 30, 25, 22, 20, 18, 17, 16, 15, 14um$, with a fixed propagating length of $L = 60um$. The fabrication process was carried on a commercially available z-cut LNOI wafer (NANOLN, Jinan Jingzheng Electronics Co., Ltd). The top LN film had a thickness of 370nm, and was bonded to a 2um silica buffer, which was deposited on a 500um LN substrate. The etching process mainly utilized electron-beam lithography (EBL). After the CAWs were fabricated, a 50-nm-thick silver film was sputtered onto the LNOI wafer using magnetron sputtering. Then, the couple-in and couple-out gratings were drilled with a focused ion beam (FIB) (FEI Dual Beam HELIOS NANOLAB 600i, 30 keV, 80 pA). At last, the samples were immersed in dilute nitric acid to remove the silver film. Detailed photographs of the sample are taken by FIB and shown in Fig.s6. The magnification in the blue box is 1200 ×, in the orange box is 1500 ×, and in the green box is 2000 ×. We cut off one waveguide to measure its cross as the red box expressed.

**Experimental set-up and Data analyzing.** We design an experimental setup, as shown in Fig.3a, to measure fixed-frequency and broadband SFG-ISU. We combine a tunable fs-laser (Spectra-physics, Mai Tai HP) and a fixed CW laser using a beam splitter (BS) and focused them onto the sample through a microscope objective (Olympus Plan Achromat Objective 20x/0.4). The polarization is controlled by a half-wave plate (HWP). We obtained the fixed-frequency SFG enhancement directly from the images captured by the sCMOS camera (Hamamatsu, ORCA-Flash 4.0, C11440-42U). To quantify the transmission of $\lambda_{SF3}$ by SFG process (set fundamental waves as $\lambda_j = 2\lambda_{SF3}$) for bandwidth broadening, we used a spectrometer (PYLON, Princeton) and the sCMOS camera. Then we normalized the SFG spectrum recorded by the spectrometer for different $\lambda_{FF1}$, and obtain the bandwidth of each CAW.

## Data Availability

The data that support the plots of this study are available. Any additional data are available from the corresponding author on reasonable request.

## Code Availability

The codes that support the findings of this study are available from the corresponding authors upon reasonable request.

## Acknowledgements

This work financially supported by the National Natural Science Foundation of China (Grant Nos. 92150302 and 92163216 to H. L., 62288101 to S. Z., 12174187 to C. S.). This work was financially supported by the National Key R&D Program of China (Grant Nos.2023YFB2805700 to C.S.). This work was sponsored by the National Research Foundation Singapore Competitive Research Program (No. NRF-CRP22-2019-0006 and NRF-CRP23-2019-0007 to Y. L.), A*STAR AME Programmatic Funds (No. A18A7b0058 to Y. L.).

## Author contributions

H.L. and Y.L. conceived the idea. Y.L. performed the theoretical derivation. C.H. conceived the calculation, analyzed the data and designed the sample. Z.Y and Y.Z fabricated the sample. H.L. and Y.Z. designed the experiments. C.H., Y.Z., X.M. and Z.L. performed the experiments. C.H., Y.L. and H.L. wrote the manuscript. All authors discussed the results. H.L. and Y.L. supervised the study.

## Competing interests

The authors declare no competing interests.

## Figures:

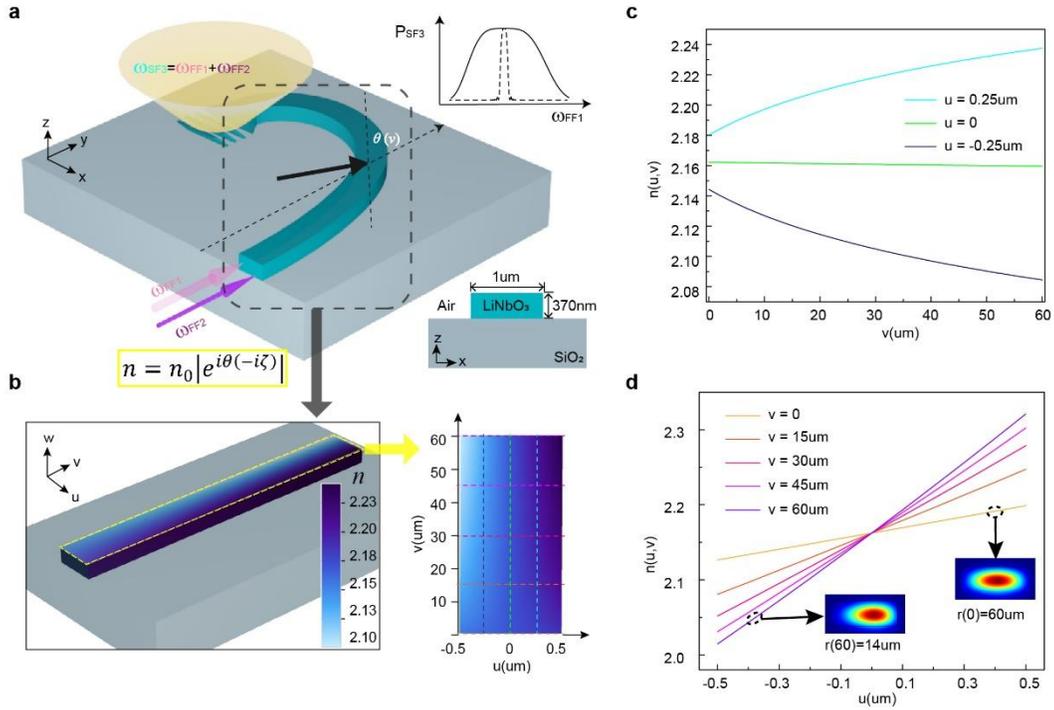

**Fig.1: Schematic of the CAW for broadband SFG-ISU**

**a,** The CAW with the spatially gradient radius $r(v)$ in the physical space $(x, y, z)$, where $r(v)$ is given by equation (2). The bottom right inset depicts the cross-section of the waveguide. The top right inset is the schematic diagram of extended bandwidth. **b,** The refractive index distribution of the transformed straight waveguide in the virtual space $(u, v, w)$. Given on the right is the 2-dimentional index map (with respect to $u$ and $v$) on the top surface of the waveguide (highlighted by the yellow box).. **c.** Refractive index distributions at $u = -0.25\text{um}, 0\text{um}, 0.25\text{um}$ (see 3 vertical dashed lines in panel b). **d.** Refractive index distributions at $v = 0\text{um}, 15\text{um}, 30\text{um}, 45\text{um}, 60\text{um}$ (see 5 horizontal dashed lines in panel b). The two insets plot the eigenmode distributions of the CAW at different $v$ ($v = 0$, $v = 60\text{um}$). The corresponding radii are $r(0) = 60\text{um}$ and $r(60um) = 14\text{um}$.

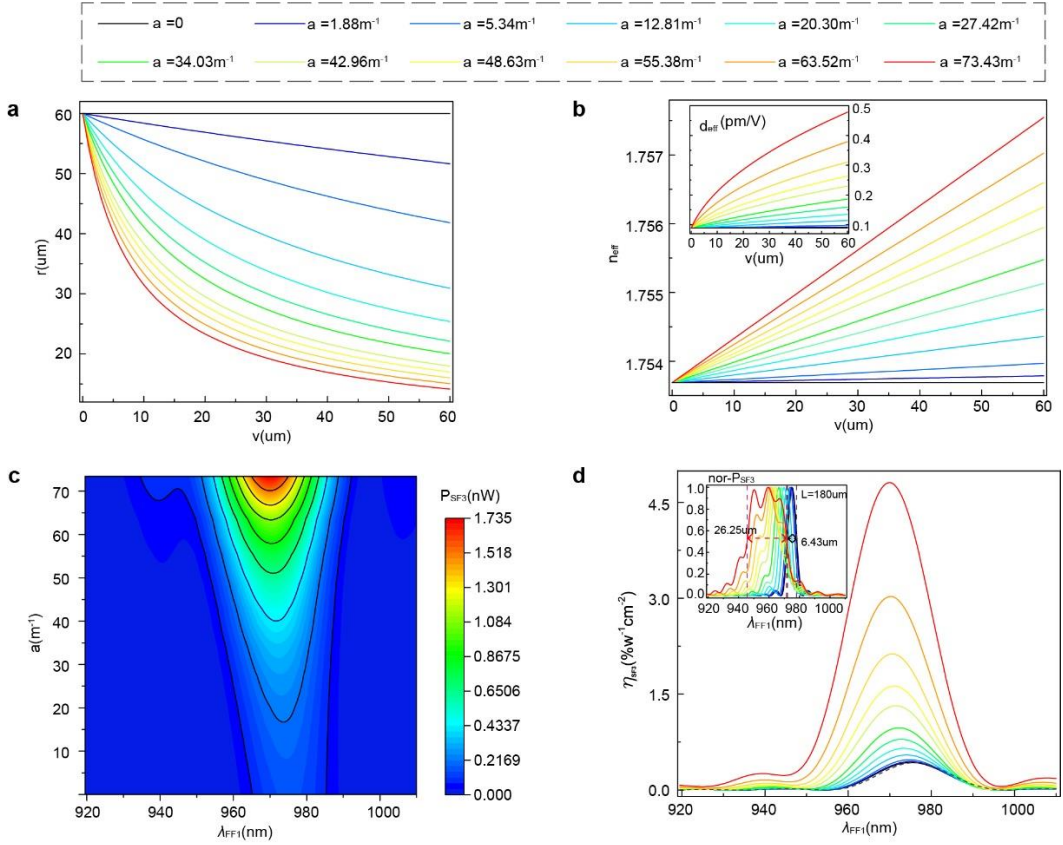

**Fig.2: Theoretical analysis of the SFG-ISU mechanism in CAWs.**
**a,** Radii of 11 **CAW**s as a function of the vertical coordinate $v$. Larger acceleration corresponds to larger gradient in $r(v)$. The legends above the figure give the exact values of accelerations of the 11 CAWs. **b,** The effective mode index distribution $n_{\text{eff}}$ at the fundamental wavelength $\lambda_{\text{FF2}} = 1064$nm for the 11 CAWs. The inset depicts the effective nonlinear coefficient $\tilde{d}_{\text{eff}}$ at $\lambda_{\text{FF1}} = 960$nm as a function of $v$. **c,** The contour plot of output sum-frequency power in terms of the fundamental wavelength $\lambda_{\text{FF1}}$ and the acceleration $a$. **d,** The normalized frequency conversion efficiency $\eta_{\text{SF3}} = P_{SF3}/(P_{FF1}P_{FF2}L^2)$ as a function of the fundamental wavelength $\lambda_{\text{FF1}}$. The black dashed curve corresponds to the straight waveguide with homogeneous refractive index. The inset demonstrates that the normalized bandwidth of the CAW at $a = 73.43 m^{-1}$ can be 4 times wider than that at $a = 0$ when the waveguide length is fixed at 180um.

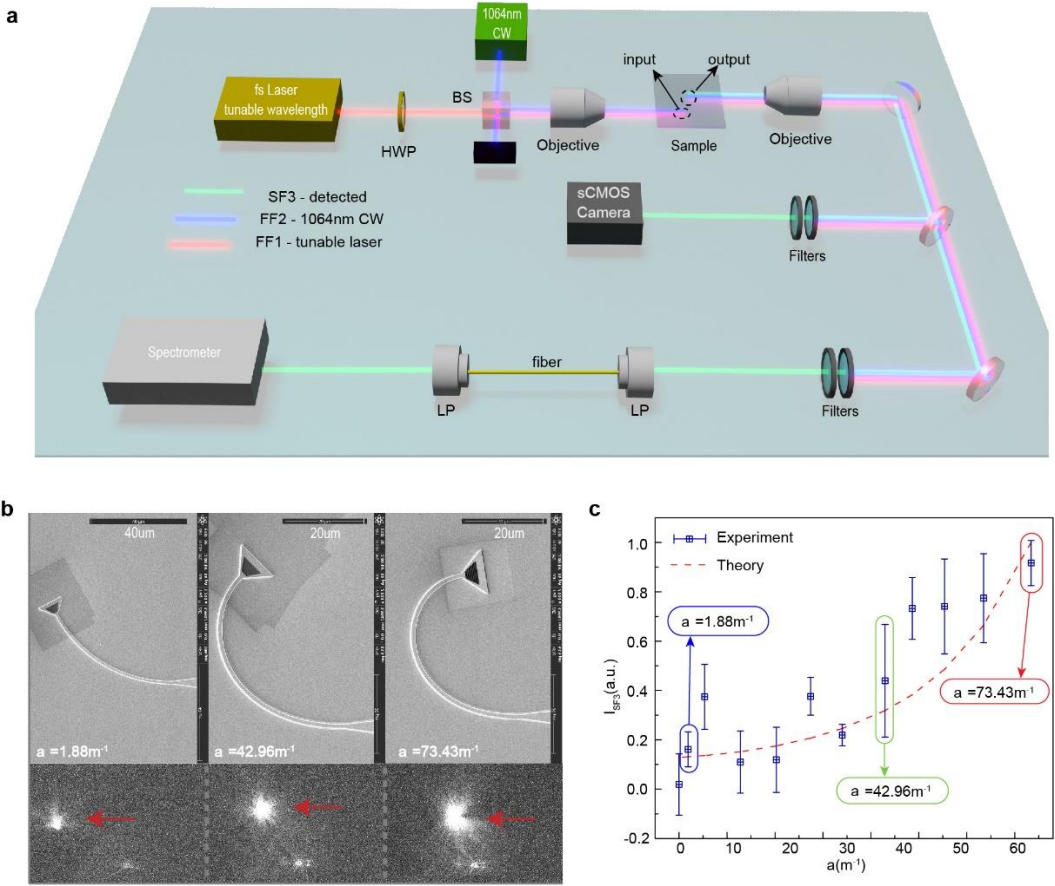

**Fig. 3: Experimental demonstration of enhanced SFG-ISU**
**a,** Schematic of the experimental setup. **b,** Bottom pictures give the SFG of CAWs taken by the sCMOS camera, where red arrows indicate the SFG output points of the CAWs. Top pictures are the corresponding SEM images of CAWs fabricated by a Focused Ion Beam (FIB). **c,** The measured SFG output intensity as a function of the acceleration $a$ (at $\lambda_{\text{FF1}} = 960nm$ ). The blue dots correspond to the experimental mean values of multiple measurements. The vertical error bars indicate the range of standard deviation (SD). The red dashed line is obtained by theoretical calculations.

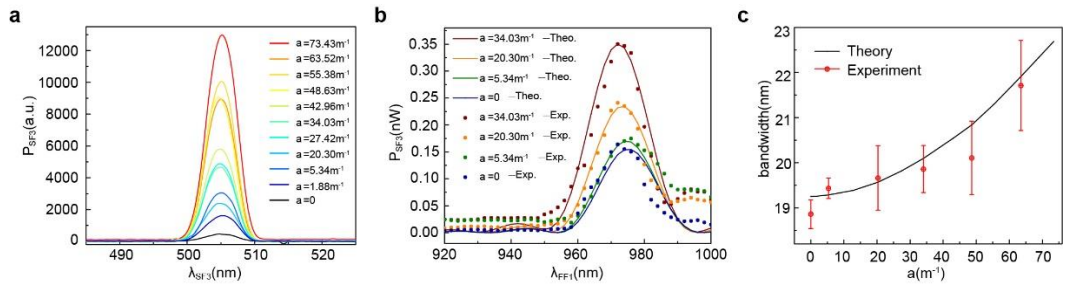

**Fig. 4: Experimental demonstration of broadband SFG in CAWs with different accelerations.** **a**, One-time tested spectrum under the same input condition with $\lambda_{\text{FF1}} = 960 nm$. The spectra are smoothened. **b,** SFG intensities of different CAWs as a function of the scanned wavelength $\lambda_{\text{FF1}}$. **c,** SFG bandwidths for CAWs with different accelerations. Red dots plot the experimental mean values. Error bars indicate the SD. Black solid line corresponds to the theoretical calculation.

# Supplementary Information

# 1. Theoretical analysis method with transformation nonlinear optics

## 1.1 Conformal Transformation from a curved waveguide in physical space to a straight one in virtual space

We consider a curved waveguide in the $\xi = x + iy$ space with any continuous bending curvature and aim to transform it into a straight waveguide in the $\zeta = u + iv$ space. To achieve this, we seek a conformal mapping $\xi = f(\zeta)$ that transforms the straight waveguide aligned along the imaginary axis (i.e. $\zeta = iv$) of the initial coordinate to a curved waveguide with a spatially gradient curvature in the physical space. We require that the conformal mapping satisfies the following conditions: 1) the conformal mapping preserves the total length of the waveguide; 2) the radius of curvature of the transformed waveguide is given $r_c(v)$, which is an arbitrary function of $v$.

We first look for the expression which describes the trajectory of the curved waveguide. Set $\theta$ as the angle between the tangential direction of the curved waveguide and the vertical axis of the physical coordinate. The conservation of the total waveguide length implies,

$$\theta(v) = \int_0^v \frac{dv'}{r_c(v')} \qquad (s1)$$

Since $dx = -\sin\theta\, dv$, $dy = \cos\theta dv$, we have $d\xi = dx + idy = ie^{i\theta(v)} dv$, assuming the transformed curved waveguide starts from $\xi = 0$ (i.e. $\zeta = 0$ is transformed to $\xi = 0$). The trajectory of the curved waveguide can then be obtained as,

$$\xi = i\int_0^v e^{i\theta(v')} dv' \qquad (s2)$$

The differentiation could be written as $\frac{d\xi}{d\zeta} = e^{i\theta(-i\zeta)}$. Thus the refractive index of the transformed waveguide is,

$$n_\zeta = n_\xi \left|\frac{d\xi}{d\zeta}\right| = n_\xi \left|e^{i\theta(-i\zeta)}\right| \qquad (s3)$$

This equation indicates that the refractive index of the transformed material $n_\zeta$ (we use $n$ instead of $n_\zeta$ to indicate the transformed refractive index in main text) is directly related to $\theta(-i\zeta)$. For the case of waveguides with a constant radius, the refractive index in the $\zeta$ space can be deduced as $n_\zeta = n_\xi e^{u/r}$, which is exactly the same as the previous work[1]. Then we present two additional cases to illustrate the universality of this conformal transformation.

**Case I:** We consider a case of a curved waveguide with a varying radius given by $r_c(v) = r_0(1 - av)$, as shown in Fig.s1(a). In this case, we obtain $\theta = \ln(1 + av)/ar$, and the corresponding conformal mapping as

$$\xi = r_0 \frac{(1 - ia\zeta)^{1 + \frac{i}{ar_0}} - 1}{1 - iar_0} \qquad (s4)$$

Then, we can deduce the refractive index of the transformed straight waveguide as

$$n_\zeta = n_\xi \left|(1 - ia\zeta)^{i/ar_0}\right| \qquad (s5)$$

The transformed refractive index of $\lambda = 960nm$ is depicted in Fig.s1 (d)

**Case II:** In this case, we consider a sinusoidally curved waveguide given by $y = 18 \cdot \sin\pi x/30$ in Fig.s1 (b). The radius varies from infinity to 5um. $\theta$ As the angle between the tangential direction of the curved waveguide and the horizontal axis of the physical coordinate, $\theta$ is defined by the function,

$$\theta(x) = \int_0^x \frac{|y''|}{(1+(y')^2)^{3/2}}\,dt \qquad (s6)$$

Here, $y'$ and $y''$ represent the first and second differential of $y$. The radius in equation (s6) is depicted in Fig.s1(b). We set $dx = \cos\theta\,dv$, $dy = \sin\theta dv$, then $dz = dx + idy = e^{i\theta}dv$, with $v$ representing the arc length. Thus we can determine the value of $v$ based on the following equation:

$$v(x) = \int_0^x \sqrt{1+(y')^2}\,dt \qquad (s7)$$

From this equation, we can get the relation between $x$ and $v$, which allows us to determine $x(v)$, $r(v)$, and $\theta(v)$. The transformed refractive index is shown in Fig.s1 (e).

Regarding the CAW in the main text, we present the transformed index in Fig.s1(f) with the radius varying as in Fig.s1(c). Thus, the conformal mapping in Eq. (s3) can transform many curved waveguides

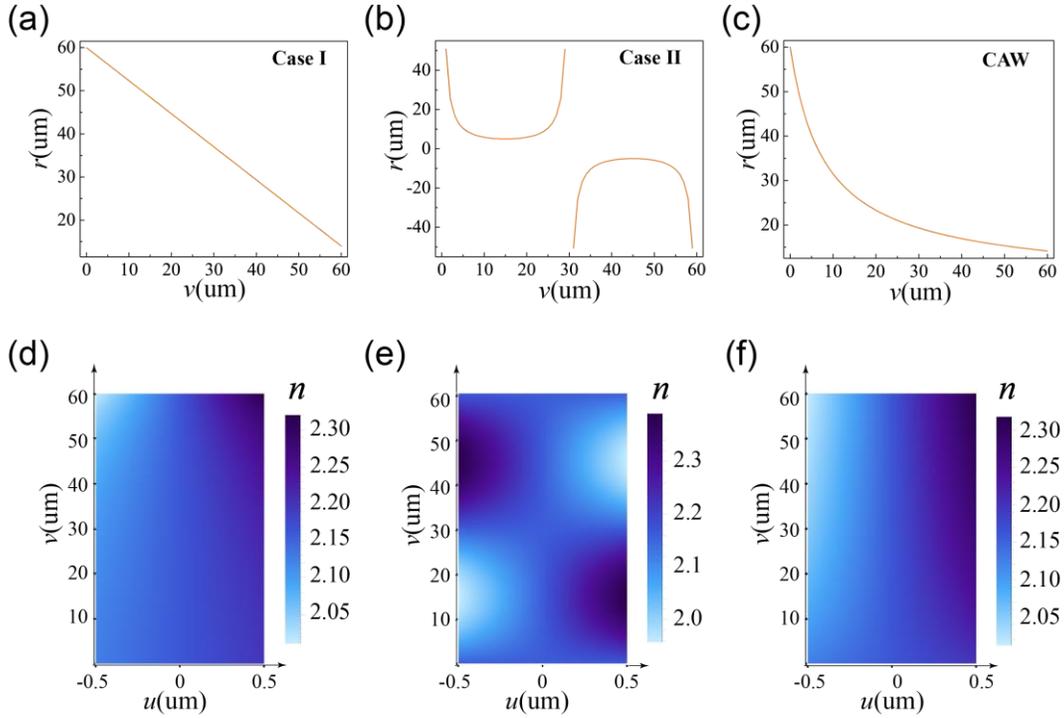

**Fig.s1 (a)** The radius varies as **Case I**. **(b)** The radius varies as **Case II**. **(c)** The radius varies as a **CAW** in main text. (d)(e)(f) are the corresponding refractive index with $\lambda = 960nm$ in **Case I**, **Case II** and the **CAW**.

into straight ones with different index distribution.

## 1.2 The propagating and the SFG process in CAWs

We define the electric field in a waveguide under the coordinate system $(u, v, w)$ as $\boldsymbol{E}(u, v, w, t) = A(v)\boldsymbol{F}_T(u, w)e^{i(kv-\omega t)}$, where the $v$-axis component is expressed as $E_v(v, t) = A(v)e^{i(kv-\omega t)}$ ($A(v)$ is the electric field amplitude along the $v$ direction, and we write $E_v$ as $E$ in main text. The transverse component is denoted by $\boldsymbol{F}_T(u, w)$. The magnitude of the wave vector $k$ is defined by the effective mode index $k = 2\pi n_{\text{eff}}/\lambda$, where $n_{\text{eff}}$ is the efficient mode refractive index and $\lambda$ is the wavelength.

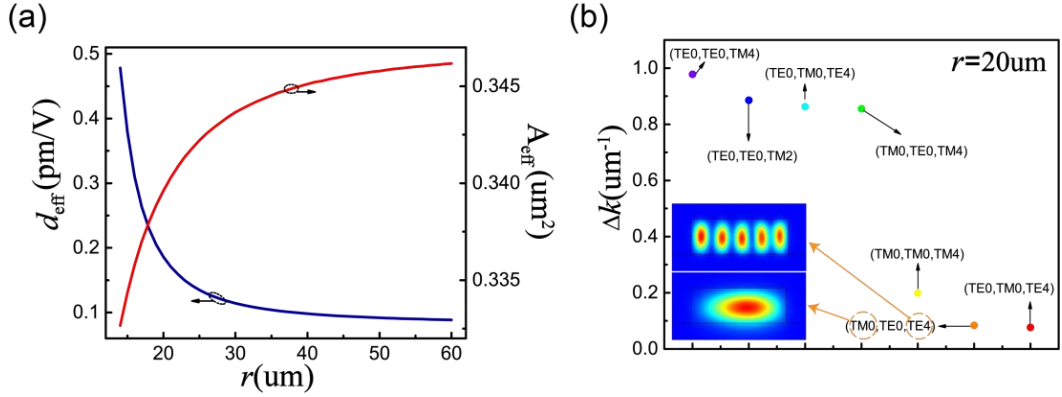

**Fig.s2 (a)** The effective nonlinear coefficient and effective mode area in a CAW. **(b)** Phase mismatches for different mode combinations at the radius of $20um$.

We especially consider the propagation in CAWs as eq. (2) expressed, for the reason of the effective mode index can be calculated as a simple function $n_{\text{eff}}(v,\omega) = n_{\text{eff}}(0,\omega)\sqrt{1-a(\omega)v}$, where $n_{\text{eff}}(0,\omega)$ is the initial effective refractive index and $a(\omega)$ denotes the acceleration of frequency $\omega$. The propagation equation in CAWs is,

$$\nabla^2 E_v - \frac{n_{\text{eff}}^2(0,\omega)(1-av)}{c^2}\frac{\partial^2 E_v}{\partial t^2} = 0 \tag{s8}$$

The propagation solution of Eq. (s8) takes the form of,

$$E_v = \text{Ai}\left[\left(\frac{n_{\text{eff}}(0,\omega)k_0}{2a}\right)^{3/2}(av-1)\right] + i\text{Bi}\left[\left(\frac{n_{\text{eff}}(0,\omega)k_0}{2a}\right)^{3/2}(av-1)\right] \tag{s9}$$

where $\text{Ai}(\cdot)$ and $\text{Bi}(\cdot)$ are the Airy functions of the first and the second kind, respectively. Since we are interested in the case where the refractive index $n_{\text{eff}}(v,\omega)$ slowly varies along $v$, $a$ must be a small quantity. Eq. (s9) reduces to

$$E_v \sim \frac{1}{(1-av)^{1/4}}\exp(i\phi) \tag{s10}$$

where $\phi = n_{\text{eff}}(0,\omega)k_0(v - av^2/4)$ denotes the phase and $k_0 = \omega/c$ is the wave vector in the vacuum.

## 1.3 The nonlinear SFG process in CAWs

We take the SFG process in CAWs to realize infrared signal upconversion as an example, i.e. $\omega_{\text{FF1}} + \omega_{\text{FF2}} = \omega_{\text{SF3}}$. We assume amplitudes of fundamental frequencies, and the sum-frequency are $A_{\text{FF1}}$, $A_{\text{FF1}}$ and $A_{\text{SF3}}$, respectively. We substitute equation (s10) to nonlinear polarization $\boldsymbol{P} = \epsilon_0\chi^{(2)}\boldsymbol{E}^2$, and deriving the nonlinear SFG coupling equation as follows:

$$\begin{aligned}\frac{dA_{\text{FF1}}}{dv} &= \kappa_{13}A_{\text{FF2}}^*A_{\text{SF3}}e^{-i\Delta\phi} \\ \frac{dA_{\text{SF3}}}{dv} &= \kappa_{31}A_{\text{FF2}}A_{\text{FF1}}e^{i\Delta\phi}\end{aligned} \tag{s11}$$

where $\Delta\phi = \phi_{\text{SF3}} - \phi_{\text{FF2}} - \phi_{\text{FF1}}$ is the phase difference. Couple-wave coefficients $\kappa_{31}$ and $\kappa_{13}$ of CAWs are functions of accelerations $a_{FF1}$, $a_{FF2}$, $a_{SF3}$, effective nonlinear coefficient $\tilde{d}_{eff}$, and coordinate $v$, which can be expressed as,

$$\kappa_{13} = \frac{-8\pi^2(1-a_{FF1}v)^{\frac{5}{4}}\tilde{d}_{eff}}{\left[\frac{\lambda_{FF1}^2}{4}a_{FF1} + 2\pi i \lambda_{FF1} n_{FF1}(2 - 3a_{FF1}v + a_{FF1}^2 v^2)\right](1-a_{FF2}v)^{\frac{1}{4}}(1-a_{SF3}v)^{\frac{1}{4}}}$$

$$\kappa_{31} = \frac{-8\pi^2(1-a_{SF3}v)^{\frac{5}{4}}\tilde{d}_{eff}}{\left[\frac{\lambda_{SF3}^2}{4}a_{SF3} + 2\pi i \lambda_{SF3} n_{SF3}(2 - 3a_{SF3}v + a_{SF3}^2 v^2)\right](1-a_{FF2}v)^{\frac{1}{4}}(1-a_{FF1}v)^{\frac{1}{4}}} \quad (s12)$$

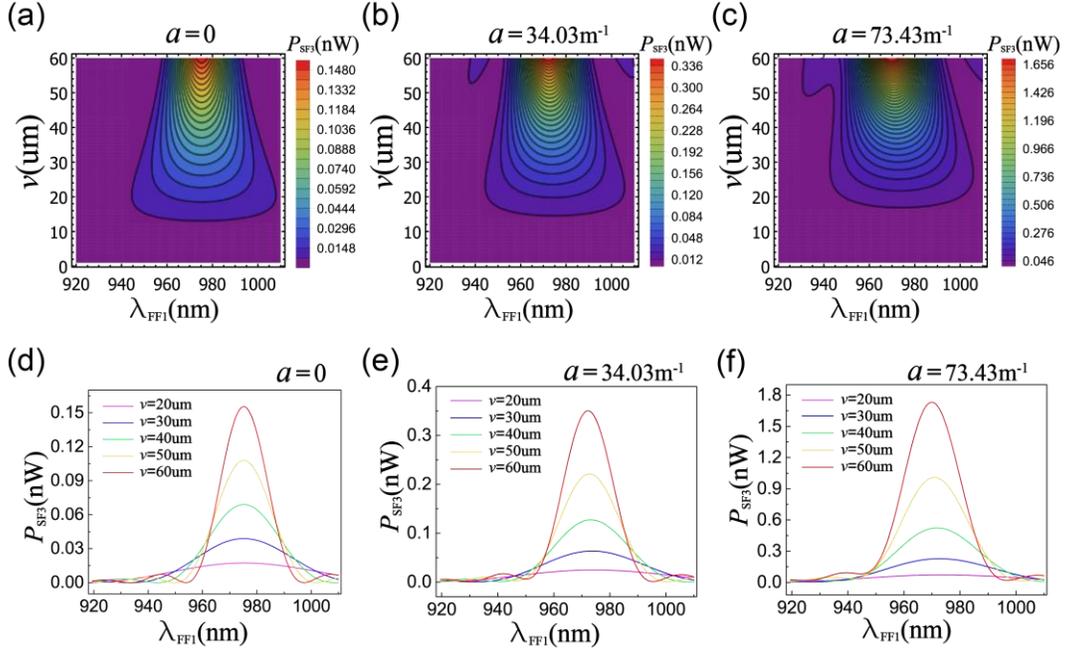

**Fig.s3** The SFG efficiency for a range of $\lambda_{FF1}$ along $v$ for different accelerations of **(a)** $a = 0$, **(b)** $a = 34.03 m^{-1}$, **(c)** $a = 73.43 m^{-1}$. While the lower part of **(e)**, **(d)**, **(f)** are 1D lines plotted in $v =$20um,30um,40um,50um,60um that corresponding to the upper part.

The effective nonlinear coefficients $\tilde{d}_{eff}$ is defined as a tensor product form $\tilde{d}_{eff} =$

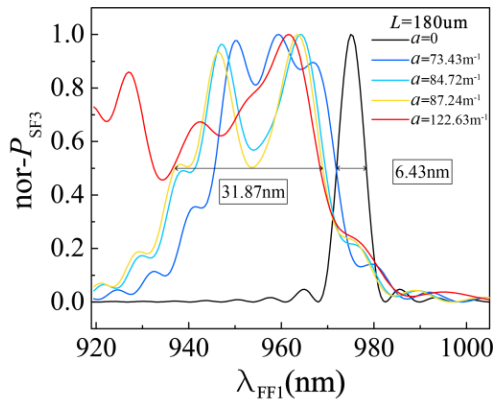

**Fig.s4** The SFG Power for a range of $\lambda_{FF1}$ with $L = 180um$ for different accelerations. Three additional accelerations of $a = 84.72 m^{-1}$, $a = 87.24 m^{-1}$ and $a = 122.63 m^{-1}$ are investigated as compared to the inset of Fig.2d. We especially remark the bandwidth of $a = 87.24 m^{-1}$ is $31.87 nm$.

$\left|\int dxdy\, \boldsymbol{F}_{T-SF3} : \overleftrightarrow{d} \cdot \boldsymbol{F}_{T-FF1} \boldsymbol{F}^*_{T-FF2}\right|$, where $\boldsymbol{F}_{T-SF3}$, $\boldsymbol{F}_{T-SF3}$ and $\boldsymbol{F}_{T-SF3}$ are transverse components of each frequency and $\overleftrightarrow{d}$ is the nonlinear $3 \times 6$ tensor matrix of crystal LN corresponding to $\chi^{(2)}$.

For the waveguide in main text, we have chosen $(TM_0, TE_0, TE_4)$ modes for FF1, FF2 and SF3 frequencies, considering the relative long coherent length. In Fig.s2 (a), we have plotted the effective nonlinear coefficient of this mode combination and the effective mode area of $TM_0$ at $\lambda = 1064nm$. The corresponding effective nonlinear coefficient related to mode overlap increases with the radius, while the effective mode area related to field intensity decreases with the radius. In Fig.s2 (b), we have depicted the phase-mismatch for different modes with a radius of $20um$, $(i,j,k)$ around points denote waveguide modes of $\omega_{FF1}$, $\omega_{FF2}$, and $\omega_{SF3}$ respectively.

We have numerically solved the CWEs in Eq. (s11) to obtain the relation between the SFG efficiency

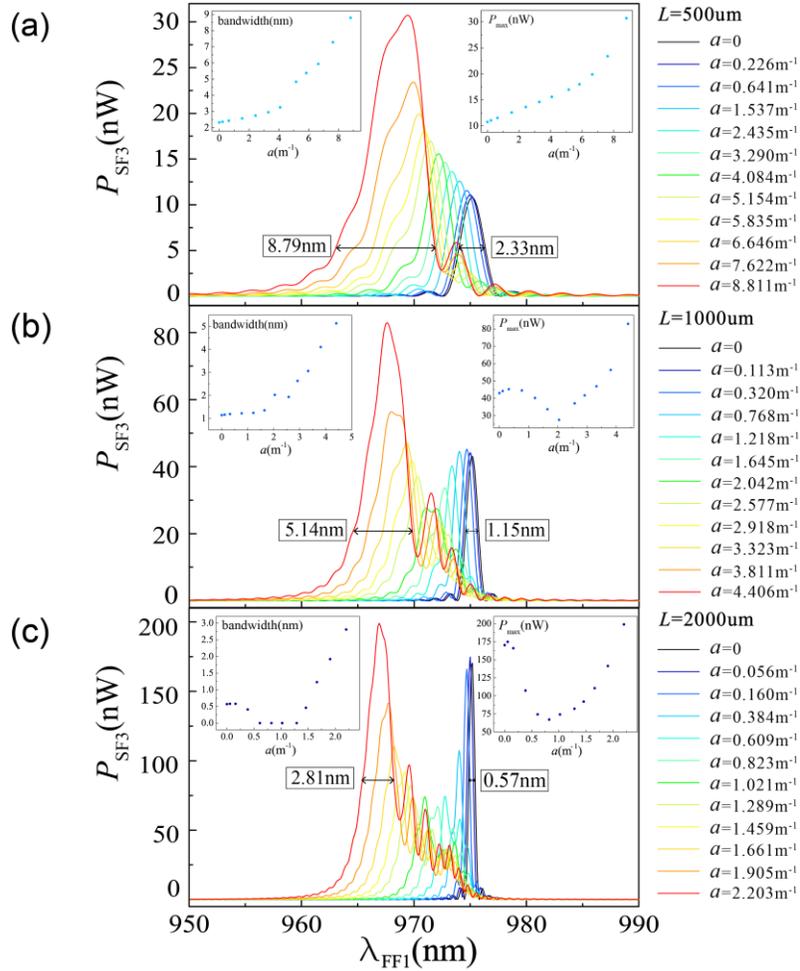

**Fig.s5** The sum-frequency power plotted as a function of $\lambda_{FF1}$ for different accelerated CAWs. (a), (b) and (c) are SFG power for three designs with $L = 500um$, $L = 1000um$ and $L = 2000um$ respectively. Each accelerations are expressed by different colors, which are listed in the right side of the figure. The left insets of each figure show the bandwidths respecting to the acceleration. The right insets are the maximum SFG power relating to accelerations. The broadening efficiency can be seen from the maximum and minimum bandwidth in black boxes.

and the propagation length $v$, and tunable fundamental wavelength for any CAWs. The intensity of SF3

can be calculated by $I_{SF3}(v) = 2n_0\epsilon_0 c|A_{SF3}(v)|^2$. The power is presented by $P_{SF3} = I_{SF3} \cdot A_{eff}$ in waveguides.

We take three examples of the extension of bandwidth caused by the acceleration $a = 0, 34.03m^{-1}, 73.43m^{-1}$, as shown in Fig.s3. In Fig.s3 (a) and (d), the bandwidth of SFG narrowed down rapidly. While in Fig.s3 (b), (c), (e) and (f), the acceleration increases the bandwidth in same $v$.

## 1.4 The bandwidth limitation in CAWs

In this section, we discuss the bandwidth limitation by manipulating the acceleration in theory. The inset of Fig.2d demonstrated that when the length is extended to $180um$, the bandwidth of $a = 73.43m^{-1}$ could be extend up to 4 times. To further analyze the normalized SFG power bandwidth, we increased the acceleration to $a = 84.72m^{-1}$, $a = 87.24m^{-1}$ and $a = 122.63m^{-1}$ as shown in Fig.s4. The results indicate that the bandwidth broadens with acceleration, where the bandwidths of CAWs with $a = 84.72m^{-1}$ and $a = 87.24m^{-1}$ are $28.22nm$ and $31.87nm$, respectively. As for $a = 122.63m^{-1}$, the bandwidth is extensively increased.

However, due to the cut-off frequency in waveguides and the limited fabrication technology, the radius cannot be further reduced to increase the acceleration. We set $r = 14um$ as the minimum radius in experiments. To demonstrate the limitation of maximum bandwidth in experiments, we compare bandwidth of CAWs under same propagating length in Fig. s5. The figure shows the sum-frequency power plotted as a function of $\lambda_{FF1}$ for three different CAW designs with L=500um, L=1000um and L=2000um, represented by (a), (b) and (c) respectively. The different colors in the figure denote the amount of acceleration achieved by each design. The left inset in each figure expresses the continuous bandwidth related to accelerations. The right inset in each figure shows the maximum frequency conversion power related to accelerations. The black boxes highlight the variation in bandwidth efficiency between the designs, as seen by the maximum and minimum bandwidths.

We observed that the bandwidth and the maximum increase with the acceleration in $L = 500um$. While for $L = 1000um$, the maximum will not increases monotonically with the acceleration. This is because of that the conversion efficiency of phase-matching gradually exceeds the gain brought by the nonlinear coefficient boosting. And for $L = 2000um$, both bandwidth and the maximum not increase monotonically with the acceleration. Therefore, we can foresee that when the conversion efficiency of phase-matching far exceeds CAWs, the advantage of broadband will not be reflect.

## 2. Experimental set-up for nonlinear optics measurement and data processing

## 2.1 Sample fabrication

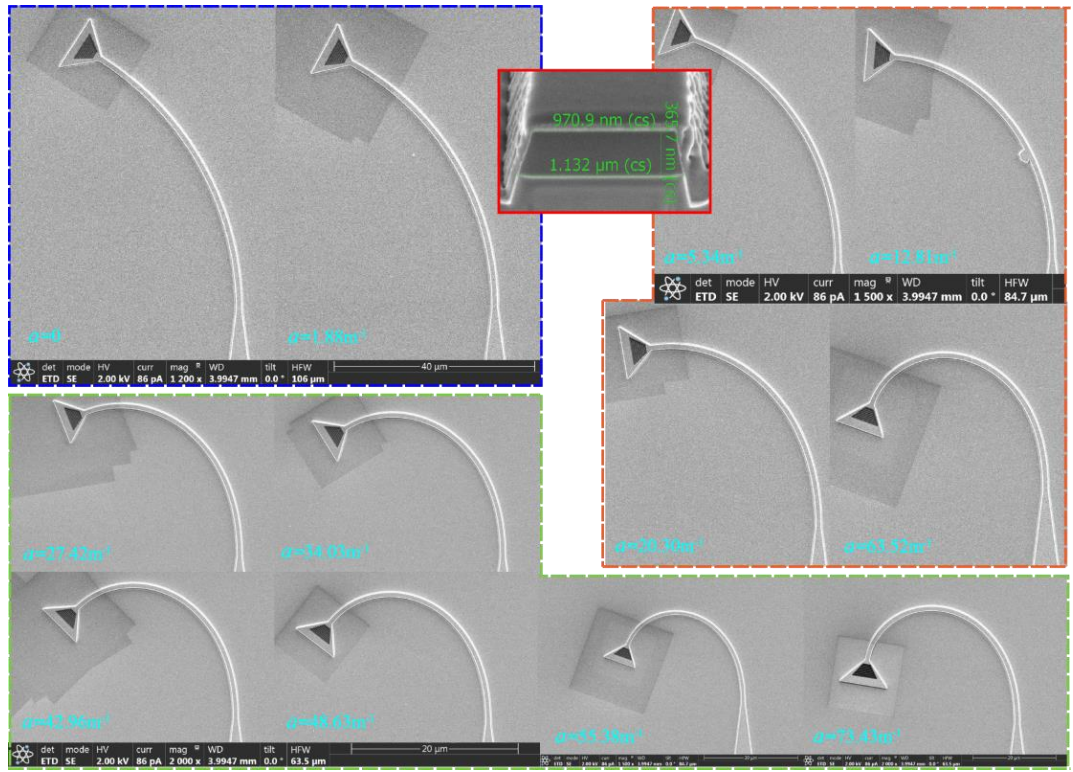

**Fig.s6** Each CAW's picture is captured by FIB. The magnification in the blue box is 1200 ×, in the orange box is 1500 ×, and in the green box is 2000 ×. The red box is the cross-section of the cut waveguide.

We fabricated all samples on LNOI, with the same propagating length $L = 60um$. In Fig.s6, we display photographs of the sample taken by a focused ion beam (FIB, FEI Dual Beam HELIOS NANOLAB). The magnification of photographs in the blue box is 1200 ×, in the orange box is 1500 ×, and in the green box is 2000 ×. We cut off one waveguide to measure its cross as the red box expressed.

## 2.2 Fixed-frequency conversion measurement.

To obtain the intensity comparison of a single frequency for $\lambda_{\mathrm{FF1}} = 960nm (\lambda_{\mathrm{FF2}} = 1064nm, \lambda_{\mathrm{SF3}} = 504.66nm)$, we collect the SFG signal by an sCMOS camera through a microscope objective (Zeiss Epiplan 40x/0.4). We add two 500nm $\pm$ 10nm filters (FB500-10) to filter out noise, as shown by the bright green dashed box in Fig. s7. We also eliminate the noise from the second-harmonic generation (SHG) process of each fundamental frequency, by making sure that the camera no longer capture any signal when block one source. The captured intensity changing by the acceleration is shown in the GIF, the corresponding CAW is also given in the inset. In another way, we also use a spectrometer (PYLON, Princeton) to test. Through a dual-lens imaging system, collimated is directly focusing into the slit of the spectrometer. In this process, we use a small aperture selection to block the incident light spot, so that the output signal directly focus into the spectrometer. Both methods can be used to test the enhancement of fixed-frequency SFG. They prove the theoretical prediction of the increasing of SFG

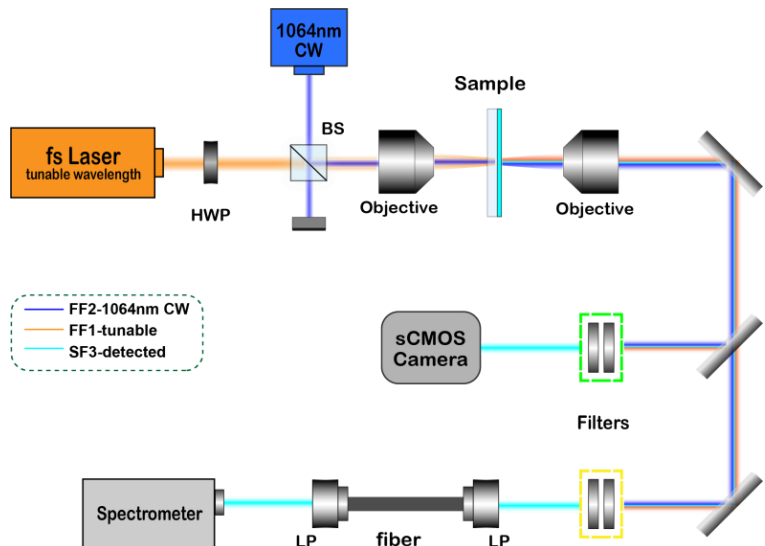

**Fig.s7** The experiment setup for SFG measurement for single frequency conversion and broadband frequency conversion

conversion efficiencies as the acceleration increases as Fig. 3c.

## 2.3 Broadband frequency conversion measurement.

We scan $\lambda_{\mathrm{FF1}}$ from $920nm$ to $1010nm$ by tuning the femtosecond laser to measure the spectrum of SFG as predicted in Fig. 2(d). There, the filter system consists of two 550nm short-pass filters (FESH0550) as the bright yellow dashed box shows. We measure the output signal by the spectrometer to distinguish the intensity of the desired $\lambda_{\mathrm{SF3}}$. Consider the coupling efficiency of gratings and transmission of detection system for different wavelength. We quqntify the coupling efficiency of the fundamental frequency by altering the fundamental wavelength under the same incident conditions.

We intruduce a fiber-lens system (FLS) to quantify the transmission of $\lambda_{\text{SF3}}$ by SHG process (set the fs laser as $\lambda_j = 2\lambda_{\text{SF3}}$), which is defined as the ratio of the intensity captured by the sCMOS camera to the intensity captured by the spectrometer. Here, the spatial signals come out from the waveguide, are focused into a fiber by a cage-structure lens, and then they output into another cage-structure lens and focusing collimated into the spectrometer. Thus the FLS also ensures that the output signals of different CAWs can be perpendicularly incident into the spectrometer, regardless of the shape of the waveguide.

## 3. Analysis of nonlinear nanofocusing

Considering the efficient field area as a function of radius in Fig.s2 (a), the field intensity in CAWs will increase with the radius decreasing. For the reason of the intensity changes as the effective mode area changes in waveguides.

$$I = \frac{P_0}{A_{\text{eff}}^L(v)} = \frac{A_{\text{eff}}^L(0)}{A_{\text{eff}}^L(v)} I_0 = \alpha_L(v) I_0 \tag{s13}$$

Where $P_0$ is the input power, $A_{\text{eff}}^L$ is the effective mode area in linear optics and $\alpha_L(v)$ is the linear enhancement parameter. As $\alpha_L(v)$ grows, the intensity increases.

This phenomenon is commonly referred to as nanofocusing[2], which is the process of increasing the electric field intensity in a structure with a gradually decreasing mode cross-section. This process has been utilized to enhance surface plasmon polaritons (SPPs)[3], including nonlinear effects enhancement[4] or plasmonic sensing[5]. Moreover, TO can be employed to achieve effective SPPs nanofocusing[6]. Through the analysis of curved waveguides, we found that the bending dimension can be utilized to compress the mode field when on-chip integration requires nanofocusing enhancement on a small footprint.

The nanofocusing of a curved waveguide with unchanged cross section normally owing to the fraction $\eta_j$ of modal power in the waveguide, which is defined as the ratio of power flow within the waveguide to total power flow of the mode[7].

$$\eta_j = \frac{\frac{1}{2}\iint_{A_{wg}} \boldsymbol{E}_j \times \boldsymbol{H}_j^* \cdot \hat{\boldsymbol{v}} dA}{\frac{1}{2}\iint_{A_\infty} \boldsymbol{E}_j \times \boldsymbol{H}_j^* \cdot \hat{\boldsymbol{v}} dA} \tag{s14}$$

Where $j$ denotes $j$th eigenmode of the waveguide, and $\hat{\boldsymbol{v}}$ is the unit vector of the propagation direction, $A_\infty$ and $A_{wg}$ denote areas of all space and just the waveguide respectively. The Eq.(s14) represents the proportion of energy involved in the nonlinear interaction process. Therefore, the field intensity act on $\chi^{(2)}$ can be defined by power per unit area determines the intensity of the interaction.

$$I_{NL} = \frac{P_{NL}}{a_{NL}} = P \times \frac{\eta_j}{a_{NL}} = P/A_{\text{eff}} \tag{s15}$$

Where $P_{NL}$ is the power involved in nonlinear interactions and $P$ is total power, $A_{\text{eff}} = a_{\text{NL}}/\eta_j$ is the effective mode area, $a_{NL}$ is the effective mode area of a given nonlinear waveguide.

$$a_{\text{NL}} = \frac{\left(\iint_{A_\infty} \boldsymbol{E}_j \times \boldsymbol{H}_j^* \cdot \hat{\boldsymbol{v}} dA\right)^2}{\iint_{A_\infty} \left(\boldsymbol{E}_j \times \boldsymbol{H}_j^* \cdot \hat{\boldsymbol{v}}\right)^2 dA} \tag{s16}$$

Therefore, nonlinear intensity $I_{NL}$ is defined as total power $P$ divided by $A_{\text{eff}}$. The nonlinear frequency conversion efficiency $\eta_{\text{NL}}$ is defined as:

$$\eta_{\text{NL}} = \frac{P_{\text{SF3}}}{P_{\text{FF2}} P_{\text{FF1}}} = \frac{I_{NL-\text{SF3}} A_{\text{eff}-\text{SF3}}}{I_{NL-\text{FF2}} A_{\text{eff}-\text{FF2}} I_{NL-\text{FF1}} A_{\text{eff}-\text{FF1}}} \tag{s17}$$

where $I_{NL-\text{SF3}}$, $I_{NL-\text{FF1}}$, $I_{NL-\text{FF2}}$ and $A_{\text{eff}-s\text{SF3}}$, $A_{\text{eff}-\text{FF1}}$, $A_{\text{eff}-\text{FF2}}$ represents the intensity and nonlinear effective mode areas for each frequency. These three modes areas influence the nonlinear

nanofocusing simultaneously. We define the enhancement by nanofocusing by:

$$\alpha(v) = \frac{A_{\text{eff}-s\text{SF3}}(v)}{A_{\text{eff}-p}(v)A_{\text{eff}-\text{FF1}}(v)} \tag{s18}$$

Then the normalized enhancement of frequency conversion efficiency in a finite length $\alpha(L)$ can be calculated, here we set $\alpha(0) = 1$ ($L = 60um$). The enhancement for different $r(L)$ are shown in Fig.s8 (a). The nanofocusing in different accelerations is shown in Fig. s8 (b)

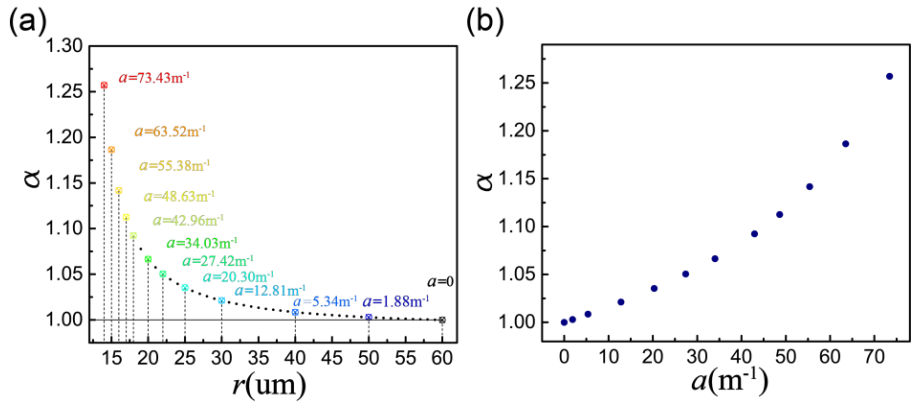

**Fig. s8** Nanofocusing effect of CAWs **(a)** The nanofocusing enhancement of different CAWs for different accelerations. **(b)** The nanofocusing enhancement related to accelerations.